\documentstyle[aps,prd,multicol]{revtex}
\tighten

\def\Lrule{\vspace*{-0.2in}\noindent\vrule width3.4in height.2pt
  depth.2pt \vrule depth0em height.5em}
\def\Rrule{\vspace{-0.1in}\hfill\vrule depth.5em height0pt \vrule
  width3.4in height.2pt depth.2pt\vspace*{-0.1in}}

\begin{document}

\title{A model for time-dependent cosmological constant}

\author{M. Novello$^{1a}$, J. Barcelos-Neto$^{2b}$,
and J.M. Salim$^1$}

\address{\mbox{}\\
$^1$ Centro Brasileiro de Pesquisas F\'{\i}sicas,\\
Rua Dr.\ Xavier Sigaud 150, Urca 22290-180 Rio de Janeiro, RJ --
Brazil\\
\mbox{}\\
$^2$ Instituto de F\'{\i}sica\\
Universidade Federal do Rio de Janeiro, RJ 21945-970 -- Brazil}
\date{\today}

\maketitle
\begin{abstract}
\hfill{\small\bf Abstract\hspace*{1.7em}}\hfill\smallskip
\par
\noindent
We present a model for a spacetime dependent cosmological {\it
constant}. We make a realization of this model based on a possible
quantum aspects of the initial stage of the universe and relate the
cosmological constant with the chiral anomaly.

\end{abstract}

\pacs{Key-words: Effective electrodynamics; non-singular cosmology}
\smallskip\mbox{}

\begin{multicols}{2}
\section{Introduction}

As Weinberg states very convincingly: ``We want to explain why the
effective cosmological {\it constant} is small now, {\it not} why it
was always small." In other words, we are looking for a dynamical
treatment of $\Lambda$.

\medskip
In the last years a proliferous series of papers dealing with a time
dependent cosmological constant appeared%
\footnotemark.
The main reason for such interest is related to the possibility to
treat the vacuum dynamically, as it was required for many inflationary
scenarios. It is not surprising that it has revealed a simpler task to
provide a model for a varying $\Lambda$ than it was for the case of a
constant one%
\footnotemark.
A certain number of different models appeared. Almost all of them deal
with  scalar-tensor theory of gravity. We can quote, for instance
\cite{diaz} and references therein. We also mention that since the
paper by \"Ozer-Taha \cite{Ozer}, many authors have been using this
spacetime dependence of $\Lambda$ with the purpose of eliminating some
difficulties from the standard cosmological model, mainly those
related to the flatness problem.

\medskip
In the present paper we propose a distinct model for a varying
$\Lambda$, based on a effective Lagrangian depending on a gauge
invariant quantity, that is constructed in terms of a field strength
and its dual. After that, we try to  associate this mechanism to a
possible quantum behavior of the initial stage of the universe. Let us
remark that nowadays we are almost all convinced that the evolution of
the Universe has been proceeded most classically. Some quantum effects
might take place at black holes \cite{hawking}, but there is no
evidence of such phenomena till now. It is then widely believed that
cosmology is essentially a classical subject. However, in the
beginning stage of the expansion of the Universe evolution, where the
matter content was formed just by elementary particles, probably
quarks and leptons, immerse in a very hot environment, it was much
more quantum. We are going to show that the chiral anomaly Lagrangian
can be conveniently adapted in order to generate a spacetime dependent
cosmological {\it constant}. This procedure would also naturally
explain why the cosmological constant is small now. Its most relevant
value would be related with the period that the universe evolved
quantically.

\medskip
The problem is that we do not know yet a consistent and complete
quantization of gravitation. A way of circumventing this problem is by
means of a semiclassical treatment, where just matter fields are
quantized and propagate in a classical curved background
\cite{birrell}. For our particular purposes in the present work, we
consider that this background also contains classical gauge fields.

\medskip
Our paper is organized as follow. In Sec. II we present the general
model to obtain a spacetime dependent cosmological {\it constant}. In
Sec. III we make a realization of this model by means of a
semiclassical treatment of matter fields interacting with a classical
curved background also containing gauge fields. We left Sec. IV for
concluding remarks and introduce an Appendix to give some details of
the calculation of the chiral anomaly in curved spacetime by using the
zeta function regularization.

\section{The model}
\renewcommand{\theequation}{2.\arabic{equation}}
\setcounter{equation}{0}

We introduce in this section a general model that can generate a time
dependent cosmological {\it constant}. Consider a Lagrangian $L$ given
by

\begin{equation}
L=L_{mat}+Y(\cal G)
\label{2.1}
\end{equation}

\noindent
where $Y$ is a nonlinear function of an invariant quantity $\cal G$
which is constructed in terms of a gauge field strength
$G_{\mu\nu}\mbox{}^a$ of a given group G and its dual
$G_{\mu\nu}^\ast\mbox{}^a=\frac{1}{2}\,
\eta_{\mu\nu\rho\lambda}\,G^{\rho\lambda a}$. We are
using the following definition for $\eta_{\mu\nu\rho\lambda}$

\begin{equation}
\eta_{\mu\nu\rho\lambda}=\sqrt{-g}\,\epsilon_{\mu\nu\rho\lambda}
\label{2.2a}
\end{equation}

\noindent
Consequently,

\begin{equation}
\eta^{\mu\nu\rho\lambda}=-\frac{1}{\sqrt{-g}}
\,\epsilon^{\mu\nu\rho\lambda}
\label{2.2b}
\end{equation}

\noindent
where $\epsilon_{\mu\nu\rho\lambda}$ and
$\epsilon^{\mu\nu\rho\lambda}$ are the usual Levi-Civita tensor
densities ($\epsilon_{0123}=1$) and $g$ is the determinant of the
metric tensor.

\medskip
For our purposes here it is not necessary to specify the gauge group,
and from now on we just report to the Abelian case for simplicity. So

\begin{equation}
{\cal G}\equiv G_{\mu\nu}^\ast\,G^{\mu\nu}
\label{2.3}
\end{equation}

In order to obtain from this form of $L$ the expression of
$T_{\mu\nu}$ we have to vary the Lagrangian with respect to
$g^{\mu\nu}$. The important property that we need is

\begin{equation}
\delta\,{\cal G}=\frac{1}{2}\,{\cal G}\,
g_{\mu\nu}\,\delta g^{\mu\nu}
\label{2.4}
\end{equation}

\noindent
From this expression and using the definition of the energy-momentum
tensor in terms of variation of the background metric provided by

\begin{equation}
T_{\mu\nu}=\frac{2}{\sqrt{-g}}\,
\frac{\delta L\sqrt{-g}}{\delta g^{\mu\nu}}
\label{2.5}
\end{equation}

\noindent
we obtain

\begin{equation}
T_{\mu\nu}=T_{\mu\nu}^{(mat)}+\Lambda\,g_{\mu\nu}
\label{2.6}
\end{equation}

\noindent
in which

\begin{equation}
\Lambda=\frac{dY}{d{\cal G}}\,{\cal G}-Y
\label{2.7}
\end{equation}

From this expression it follows that the effective Lagrangian
contributes to the energy-momentum tensor with a term that
is proportional to the metric tensor of the background geometry. This
is interpreted, in the Einstein General Relativity theory as a
spacetime dependent cosmological {\it constant}. We observe that if
$Y$ is linear in $\cal G$, there is no contribution for $\Lambda$.

\medskip
It is opportune to mention that besides the ordinary interaction of
matter with the gauge field $A$, through a conserving current $J^{\mu}
_{A}$, the field can interact with axionic matter $a$ via a Lagrangian
that depends on the product $a\,{\cal G}$ \cite{Kolb}

\begin{equation}
L_{int} = L(a\,{\cal G})
\label{2.8}
\end{equation}

\noindent
The presence of this interaction will not affect the general lines of
the mechanism to obtain the spacetime dependent cosmological {\it
constant} but, of course, it will modify its value %
\footnotemark.

\section{A quantum origin for $\Lambda$}
\renewcommand{\theequation}{3.\arabic{equation}}
\setcounter{equation}{0}

In this section we associate the Lagrangian $Y$ to a possible quantum
scenario of the beginning universe. This would lead to a natural time
dependence of the cosmological {\it constant}, in a sense that it
would be relevant just in the period of this quantum scenario.

\medskip
Let us then consider an idealized situation with the universe being
filled with quantum (fermionic) matter fields $\psi$ interacting with
some external gauge fields $A_\mu$ and also with an strong
gravitational background%
\footnotemark.
We have the following action

\begin{equation}
S=i\int d^4x\sqrt{-g}\,
\bar\psi\gamma^\mu\bigl(\nabla_\mu-ie\,A_\mu)\psi
\label{3.1}
\end{equation}

\bigskip\noindent
where $\gamma^\mu$ are the curved-space Dirac matrices, depending on
the spacetime, that are related to a local flat-space by

\begin{equation}
\gamma^\mu(x)=e^\mu_a(x)\gamma^a
\label{3.2}
\end{equation}

\bigskip\noindent
$e^\mu_a$ are the tetrad fields and satisfy the standard relations

\begin{eqnarray}
&&e_{\mu a}e_\nu^a=g_{\mu\nu}
\nonumber\\
&&e_{\mu a}e^{\nu a}=\delta_\mu^\nu
\nonumber\\
&&e_{\mu a}e^{\mu^b}=\delta^b_a
\nonumber\\
&&e_{\mu a}e^\mu_b=\eta_{ab}
\label{3.3}
\end{eqnarray}

\bigskip\noindent
Consequently,

\begin{eqnarray}
g&=&\det\,(e^a_\mu)\det\,(e^b_\nu)\,\det\,(\eta_{ab})
\nonumber\\
&=&-\,(\det\,e)^2
\label{3.4}
\end{eqnarray}

\noindent
The covariant derivative $\nabla_\mu$ acting on the spinorial field
$\psi$ means

\begin{equation}
\nabla_\mu\psi=(\partial_\mu+\Gamma_\mu)\,\psi
\label{3.5}
\end{equation}

\bigskip\noindent
where $\Gamma_\mu(x)=\frac{1}{2}\omega_{\mu ab}\Sigma^{ab}$ is the
spin connection. Here, $\Sigma^{ab}$ are flat-space time matrices
defined by $\Sigma^{ab}=\frac{1}{4}[\gamma^a,\gamma^b]$. The
quantities $\omega_{\mu ab}$ can be written in terms of the tetrad
fields by imposing that the covariant derivative of the connection is
zero. The result is

\begin{equation}
\omega_{\mu ab}=\frac{1}{2}\Bigl[e^\nu_a\,
(\partial_\mu e_{\nu b}-\partial_\nu e_{\mu b}
+e^c_\mu e^\rho_b\partial_\rho e_{\nu c})
- (a\leftrightarrow b)\Bigr]
\label{3.6}
\end{equation}

\bigskip
We have not considered the mass of particles by virtue of the high
energies evolved in the initial stage of the universe. Since there is
no mass, the action (\ref{3.1}), besides gauge invariance, is also
invariant by chiral gauge transformation,
namely

\begin{equation}
\delta\psi(x)=-i\gamma_5\xi(x)\psi(x)
\label{3.7}
\end{equation}

\bigskip
\noindent
The $\gamma_5(x)$ matrix in curved space is the same as the usual flat
$\gamma_5$, i.e.

\begin{eqnarray}
\gamma_5(x)&=&i\sqrt{-g}\,\gamma^0(x)\gamma^1(x)\gamma^2(x)\gamma^3(x)
\nonumber\\
&=&\frac{i}{4!}\sqrt{-g}\,\epsilon_{\mu\nu\rho\lambda}
\gamma^\mu\gamma^\nu\gamma^\rho\gamma^\lambda
\nonumber\\
&=&\frac{i}{4!}\sqrt{-g}\,\epsilon_{\mu\nu\rho\lambda}
e^\mu_ae^\nu_be^\rho_ce^\lambda_d
\gamma^a\gamma^b\gamma^c\gamma^d
\nonumber\\
&=&\frac{i}{4!}\sqrt{-g}\,\epsilon_{abcd}\det\,(e^\mu_a)
\gamma^a\gamma^b\gamma^c\gamma^d
\nonumber\\
&=&\gamma_5
\label{3.8}
\end{eqnarray}

\bigskip
The chiral invariance is not maintained in the quantum counterpart,
what leads to an effective action given by (See Appendix A)

\begin{eqnarray}
S_{eff}&=&S-\frac{ie^3}{16\pi^2}\int d^4x\,\xi(x)\,
\epsilon^{\mu\nu\rho\lambda}G_{\mu\nu}G_{\rho\lambda}
\nonumber\\
&=&S+\frac{ie^3}{16\pi^2}\int d^4x\sqrt{-g}\,\xi(x)\,
\eta^{\mu\nu\rho\lambda}G_{\mu\nu}G_{\rho\lambda}
\label{3.9}
\end{eqnarray}

\bigskip\noindent
Even though the above equation contains a term that could be
identified with $\cal G$ it would not generate a spacetime dependent
cosmological {\it constant} because it is linear in $\cal G$. In order
to associate this axial anomaly with the origin of $\Lambda(x)$, we
redefine the gauge field $A_\mu$ like

\begin{equation}
A_\mu=B_\mu\,\phi
\label{3.10}
\end{equation}

\bigskip\noindent
where $\phi$ is considered to be some gauge invariant quantity.
Consequently the gauge transformation of $B_\mu$ reads

\begin{equation}
\delta B_\mu=\frac{\partial_\mu\xi}{\phi}
\label{3.11}
\end{equation}

\bigskip\noindent
Let us replace $A_\mu$ by $B_\mu\phi$ into the expression
(\ref{3.9}).

\end{multicols}
\renewcommand{\theequation}{3.\arabic{equation}}
\Lrule

\begin{eqnarray}
S_{eff}&=&S+\frac{ie^3}{4\pi^2}\int d^4x\,\sqrt{-g}\,\xi(x)\,
\eta^{\mu\nu\rho\lambda}
\partial_\mu(B_\nu\phi)\partial_\rho(B_\lambda\phi)
\nonumber\\
&=&S+\frac{ie^3}{4\pi^2}\int d^4x\,\sqrt{-g}\,\xi(x)\,
\eta^{\mu\nu\rho\lambda}
(\partial_\mu B_\nu\phi+B_\nu\partial_\mu\phi)
(\partial_\rho B_\lambda\phi+B_\lambda\partial_\rho\phi)
\nonumber\\
&=&S+\frac{ie^3}{4\pi^2}\int d^4x\,\sqrt{-g}\,\xi(x)\,
\eta^{\mu\nu\rho\lambda}
(\partial_\mu B_\nu\partial_\rho B_\lambda\phi^2
+2\partial_\mu B_\nu B_\lambda\phi\partial_\rho\phi)
\nonumber\\
&=&S-\frac{ie^3}{4\pi^2}\int d^4x\,\sqrt{-g}\,
\eta^{\mu\nu\rho\lambda}
\partial_\mu B_\nu B_\lambda\phi^2\partial_\rho\xi
\label{3.12}
\end{eqnarray}

\bigskip

\Rrule
\begin{multicols}{2}
\noindent
where we have disregarded total derivative terms inside the action. If
one identifies $\phi$ with $\cal G$, what is in agreement with
the gauge invariance condition for $\phi$, we observe that the second
term of (\ref{3.12}) will be quadratic in ${\cal G}$. However,
there is another part that is not topological invariant. Let us
convenient rewrite the expression (\ref{3.12}) as

\begin{equation}
S_{eff}=S+\int d^4x\,\sqrt{-g}\,{\cal K}\,{\cal G}^2
\label{3.13}
\end{equation}

\noindent
where ${\cal K}=-\frac{ie^3}{4\pi^2}\eta^{\mu\nu\rho\lambda}
\partial_\mu B_\nu B_\lambda\partial_\rho\xi$. Similarly to the
variation of $\cal G$ with respect to the metric tensor, we also have
for $\cal K$

\begin{equation}
\delta{\cal K}=\frac{1}{2}\,{\cal K}\,g_{\mu\nu}\delta g^{\mu\nu}
\label{3.14}
\end{equation}

\noindent
Even though the corresponding Lagrangian of Eq. (\ref{3.13}) is
slightly different of the term $Y$ of the Lagrangian
introduced in the initial model, we can show that this term also leads
to a cosmological {\it constant} $\Lambda$ given by

\begin{equation}
\Lambda=2{\cal K}{\cal G}^2
\label{3.15}
\end{equation}

\noindent
which shows that the chiral anomaly can be associated to a origin for
the time-dependent cosmological {\it constant}.

\medskip
It is opportune to remark that the above mechanism which generates the
cosmological term $\Lambda$ also requires a modification on the matter
Lagrangian such that the total energy momentum tensor $T_{\mu\nu}=
T_{\mu\nu}^{mat}+\Lambda g_{\mu\nu}$ is conserved.

\section{Conclusion}
In the first part of this work we have presented a general model to
obtain a spacetime dependent cosmological {\it constant}. After that,
we have shown that this model can be realized by means of a
semiclassical treatment of an idealized situation related to the
primordial universe, where we consider it was filled with quantum
fermionic matter interacting with a external gauge field and with a
strong classical gravitational background. Our main purpose in
associating this model to a possible quantum scenario of the
primordial universe is that the so obtained cosmological {\it
constant} would have a significant value just while the universe
evolved quantically. This naturally explains why the cosmological {\it
constant} is small nowadays, when the expansion of the universe is
widely accepted to be classical.

\medskip
The next natural step of this research is to consider this idea in
some cosmological model in order to see how the cosmological term that
comes from Eq. (\ref{3.12}) will modify the dynamics of the Einstein
equation. This is presently under study and possible results shall be
reported elsewhere \cite{Novello}.

\vspace{1cm}
\noindent {\bf Acknowledgment:} This work is supported in part by
Conselho Nacional de Desenvolvimento Cient\'{\i}fico e Tecnol\'ogico
- CNPq (Brazilian Research Agency).

\appendix
\renewcommand{\theequation}{A.\arabic{equation}}
\setcounter{equation}{0}
\section*{Appendix A}

In this Appendix, we briefly discuss the obtainment of the chiral
anomaly by using the Lagrangian path integral formalism and the zeta
function regularization, that are both consistent with the curved
space formulation.

Let us write down the vacuum functional corresponding to the action
$S$ given by (\ref{3.1}), i.e.

\begin{equation}
Z=\int[d\bar\psi][d\psi]\,e^{iS}
\label{A.1}
\end{equation}

\bigskip\noindent
The functional integrations are just over $\bar\psi$ and $\psi$
because just the fermionic fields are considered to be quantum. They
evolve in a curved classical background that also contains gauge
fields.

\medskip
We expand $\psi$ and $\bar\psi$ in terms of a complete set of
orthonormal eigenfunctions $\{\Phi_m\}$ of the operator
$D\!\!\!\!/=\gamma^\mu(\nabla_\mu-ieA_\mu)$, i.e.

\begin{equation}
D\!\!\!\!/\,\Phi_m=\gamma^\mu(\nabla_\mu-ieA_\mu)\,\Phi_m
=\lambda_m\Phi_m
\label{A.2}
\end{equation}

\noindent
Since the set is complete and orthonormal we have

\begin{eqnarray}
&&\int d^4x\,\sqrt{-g}\,\Phi_m^\dagger(x)\Phi_n(x)=\delta_{mn}
\label{A.3}\\
&&\sum_m\Phi_m(x)\Phi_m^\dagger(y)=\frac{\delta(x-y)}{\sqrt{-g}}
\label{A.4}
\end{eqnarray}

\noindent
This allows expansions of the fermion fields $\psi(x)$ and
$\bar\psi(x)$ as

\begin{eqnarray}
\psi(x)&=&\sum_ma_m\,\Phi_m(x)
\nonumber\\
\bar\psi(x)&=&\sum_m\Phi_m^\dagger(x)\,b_m
\label{A.5}
\end{eqnarray}

\noindent
so that

\begin{equation}
[d\bar\psi][d\psi]=\prod_mdb_mda_m
\label{A.6}
\end{equation}

\noindent
$a_m$ and $b_m$ are elements of a Grassmannian algebra.

\medskip
Considering the infinitesimal chiral transformations given by
(\ref{3.7}), we have

\begin{eqnarray}
\psi^\prime(x)&=&[1-ie\gamma_5\xi(x)]\,\psi(x)
\nonumber\\
\bar\psi^\prime(x)&=&\bar\psi(x)\,[1-ie\gamma_5\xi(x)]
\label{A.7}
\end{eqnarray}

\noindent
and also expanding $\psi^\prime$ and $\bar\psi^\prime$ in terms of the
complete set $\{\Phi_m\}$, we obtain

\begin{eqnarray}
\psi^\prime(x)&=&\sum_ma_m^\prime\,\Phi_m(x)
\nonumber\\
\bar\psi^\prime(x)&=&\sum_m\Phi_m^\dagger(x)\,b_m^\prime
\label{A.8}
\end{eqnarray}

\noindent
The combination of (\ref{A.5}), (\ref{A.7}), and (\ref{A.8}) leads to

\begin{eqnarray}
a_m^\prime&=&\sum_nC_{mn}\,a_n
\nonumber\\
b_m^\prime&=&\sum_nC_{mn}\,b_n
\label{A.9}
\end{eqnarray}

\noindent
where

\begin{equation}
C_{mn}=\delta_{mn}-ie\int d^4x\,\sqrt{-g}\,\xi(x)\,
\Phi_m(x)\gamma_5\Phi_n(x)
\label{A.10}
\end{equation}

\noindent
In the case of Grassmann variables, we have

\begin{eqnarray}
\prod_mda_m&=&\det(C_{mn})\,\prod_mda_m^\prime
\nonumber\\
\prod_mdb_m&=&\det(C_{mn})\,\prod_mdb_m^\prime
\label{A.11}
\end{eqnarray}

\noindent
Consequently

\begin{equation}
[d\bar\psi][d\psi]=\Bigl(\det(C_{mn})\Bigr)^2
[d\bar\psi^\prime][d\psi^\prime]
\label{A.12}
\end{equation}

\noindent
Considering the expression for the matrix $C_{mn}$ given by
(\ref{A.10}), we have

\end{multicols}
\renewcommand{\theequation}{A.\arabic{equation}}
\Lrule

\begin{eqnarray}
\det(C_{mn})&=&\det\Bigl(\delta_{mn}-ie\int d^4x\,\sqrt{-g}\,
\xi(x)\,\Phi_m^\dagger(x)\gamma_5\Phi_n(x)\Bigr)
\nonumber\\
&=&\exp{\rm tr}\ln\Bigl(\delta_{mn}-ie\int d^4x\,\sqrt{-g}\,
\xi(x)\,\Phi_m^\dagger(x)\gamma_5\Phi_n(x)\Bigr)
\nonumber\\
&\simeq&\exp{\rm tr}\Bigl(-ie\int d^4x\,\sqrt{-g}\,
\xi(x)\,\Phi_m^\dagger(x)\gamma_5\Phi_n(x)\Bigr)
\nonumber\\
&=&\exp\Bigl(-ie\int d^4x\,\sqrt{-g}\,
\xi(x)\sum_m\Phi_m^\dagger(x)\gamma_5\Phi_n(x)\Bigr)
\label{A.13}
\end{eqnarray}

\Rrule
\begin{multicols}{2}
\noindent
From (\ref{A.4}) we see that the sum which appears in the expression
above is badly divergent. We are going to regularize it by using the
zeta function technique in curved spacetime \cite{zeta}. The main
result of this procedure can be resumed in the expression
\cite{DeWitt}

\begin{equation}
\sum_m\Phi^\dagger_m(x)\gamma_5\Phi_m(x)
=\lim_{s\rightarrow0}{\rm tr}[\gamma_5\zeta(x,s)]
\label{A.14}
\end{equation}

\noindent
where $\zeta(x,s)$ is the generalized zeta function, which is related
to the usual zeta function $\zeta(x)$ by

\begin{equation}
\zeta(s)={\rm tr}\int d^4x\,\sqrt{-g}\,\zeta(x,s)
\label{A.15}
\end{equation}

\noindent
We can obtain informations of the behavior of the generalized zeta
function by means of the heat equation and one can show that the
generalized zeta function can be associated to powers of $s$ with
coefficients depending on $x$, denoted by $[a_n(x)]$. In the
particular case of the operator $/\!\!\!D^2$ we obtain the
general
result

\begin{equation}
\lim_{s\rightarrow0}\zeta(x,s)=\left\{
\begin{array}{ll}
\frac{\sqrt{-g}}{(4\pi)^{d/2}}\,[a_{d/2}]&\mbox{$d$ even}\\
0&\mbox{$d$ odd}
\end{array}
\right.
\label{A.16}
\end{equation}

\noindent
where $d$ is the spacetime dimension. So, for $d=4$, $\zeta(x,0)$ is
just related to $[a_2]$. These coefficients are already evaluated in
literature for some specific operators. For the case of the operator
$/\!\!\!D^2$, we have \cite{DeWitt}

\end{multicols}
\renewcommand{\theequation}{3.\arabic{equation}}
\Lrule

\begin{eqnarray}
[a_2]&=&-\frac{1}{120}\,R_{;\mu}\mbox{}^\mu
+\frac{1}{288}\,R^2
-\frac{1}{180}\,R_{\mu\nu}R^{\mu\nu}
+\frac{1}{180}\,R_{\mu\nu\rho\lambda}R^{\mu\nu\rho\lambda}
\nonumber\\
&&+\frac{1}{24}\,[\gamma^\mu,\gamma^\nu]G_{\mu\nu;\rho}\mbox{}^\rho
+\frac{1}{24}\,R\,[\gamma^\mu,\gamma^\nu]G_{\mu\nu}
+\frac{1}{32}\,([\gamma^\mu,\gamma^\nu]G_{\mu\nu})^2
+\frac{1}{12}G_{\mu\nu}G^{\mu\nu}
\label{A.17}
\end{eqnarray}

\Rrule
\begin{multicols}{2}
\noindent
From (\ref{A.14}) we notice that the regularization of the sum
$\sum_m\Phi_m^\dagger\gamma_5\Phi_m$ contains $\gamma_5$ times
$\zeta(x,0)$. Hence, just the term
$\frac{1}{32}([\gamma^\mu,\gamma^\nu]G_{\mu\nu})^2$ will contributes
for the sum, which leads to the anomaly term given by (\ref{3.9}).

\end{multicols}

\begin{thebibliography}{100}
\bibitem[3]{} We are going to see in the next section that the
realization of this model is achieved by a slightly more complicate
quantity than $Y({\cal G})$, but the general lines are precisely the
same presented here.
\bibitem[4]{} The interaction of the matter with the G-field is given
through a current $J_{\mu}\mbox{}^{A}$ but since it is conserved it
does not contribute to the total $T_{\mu\nu}.$
\end{thebibliography}
\end{document}